\documentclass[twocolumn,notitlepage,10pt,pre,,showpacs]{revtex4-1}
\usepackage[english]{babel}
\usepackage{amsmath}
\usepackage{amssymb}
\usepackage{bm}
\usepackage{graphicx}
\usepackage{mathptmx}
\usepackage{color}
\usepackage{float}
\usepackage{siunitx}

\DeclareMathOperator{\sgn}{sgn}

\begin{document}

\title{Exchange corrections in a low temperature plasma}
\author{Robin Ekman, Jens Zamanian and Gert Brodin}
\affiliation{Department of Physics, Ume{\aa } University, SE--901 87 Ume{\aa
}, Sweden,}

\begin{abstract}
We have studied the exchange corrections to linear electrostatic wave
propagation in a plasma using a quantum kinetic formalism. Specifically we
have considered the zero temperature limit. In order to simplify the
calculations we have focused on the long wavelength limit, i.e. wavelengths
much longer than the de Broglie wavelength. For the case of ion-acoustic
waves we have calculated the exchange correction both to the damping rate
and the real part of the frequency. For Langmuir waves the frequency shift
due to exchange effects is found. Our results are compared with the
frequency shifts deduced from commonly used exchange potentials which are
computed from density functional theory.
\end{abstract}

\maketitle

\section{Introduction}

Recently much work has been devoted to quantum plasmas, see e.g. the books
and review articles \cite%
{Haas-book,Shukla-Eliasson-RMP,Bonitz-1998,manfredi2006}. The research is
motivated by an interest in e.g. quantum wells \cite{Manfredi-quantum-well},
spintronics \cite{Spintronics}, plasmonics \cite{Atwater-Plasmonics}, laser
plasma interaction \cite{glenzer-redmer}, astrophysical applications \cite%
{Astro} or general theory development \cite{Zamanian-2010-NJP}. The
theoretical descriptions range from hydrodynamical equations (e.g. Refs. 
\cite{Haas-book, manfredi2006, Manfredi-DFT}) to quantum kinetic models
(e.g. Refs. \cite{Bonitz-1998, Zamanian-2010-NJP, Asenjo-2009,
Wigner-example}] and field theoretical approaches (e.g. Ref. \cite%
{Field-theory}). Most models include the physical effects of particle
dispersion and Fermi pressure, and in some cases the magnetic dipole force
and magnetization due to the electron spin \cite{Zamanian-2010-NJP,
Asenjo-2009,Lundin2010,Brodin-2010}. An important effect that is sometimes
accounted for (e.g. Refs. \cite%
{Zamanian-exchange,Zamanian-II-exchange,Manfredi-DFT,DFT-HF-LF,DFT-LF,DFT-rel,DFT-semicond,Andreev-exchange}%
) but often overlooked is the exchange effects resulting from the total
antisymmetry of the electron wavefunction. A popular approach to include the
effects of exchange interaction has been to apply density dependent
potentials deduced from density functional theory (DFT) \cite%
{Manfredi-DFT,DFT-HF-LF,DFT-LF,DFT-rel,DFT-semicond}. An advantage with this
is that the resulting fluid models becomes comparatively simple once the
exchange potentials are established. As a consequence problems involving
both high-frequency dynamics \cite{Manfredi-DFT,DFT-HF-LF} as well as
low-frequency (ion-acoustic) dynamics \cite{DFT-HF-LF,DFT-LF} can be
addressed in a straightforward way, also for nonlinear problems \cite%
{DFT-HF-LF}. A drawback is that the calculation of DFT potentials typically
involve approximations (e.g. the local density approximation (LDA)) whose
accuracy can be hard to estimate beforehand. Thus there is a general need to
validate results derived from DFT by independent methods.

In the present paper we calculate the exchange contribution to the
ion-acoustic dispersion relation using quantum kinetic theory derived from
first principles. Previous works along this line \cite%
{Zamanian-exchange,Zamanian-II-exchange} have assumed the ordering $T\gg
T_{F}$ (where $T$ is the temperature and $T_{F}$ is the Fermi temperature),
which has prevented a direct comparison with results based on DFT potentials
that have considered the opposite ordering. In this paper we focus on the
low-temperature limit $T\ll T_{F}$ and evaluate the exchange contribution to
the ion-acoustic dispersion relation in the Hartree-Fock approximation to
first order in perturbation theory. We deduce that the effects of the
exchange term is to increase the phase-velocity of the ion-acoustic mode and
to increase the linear damping rate (which is due to wave-particle
interaction). Moreover, as a confirmation of the correctness of the quantum
kinetic formalism, we compare results from our quantum kinetic formalism
with previous results for the exchange contribution of high-frequency
Langmuir waves. In this latter case we recover the results of Refs. \cite%
{Roos-1961,Kanazawa-1960,Nozieres-1958} exactly within the outlined
approximation scheme.

Finally we compare our findings with results based on commonly used DFT
potentials \cite{Manfredi-DFT,DFT-HF-LF,DFT-LF,DFT-semicond}. As the DFT
potentials are incorporated in a fluid formalism, no comparison can be made
for the damping due to wave-particle interaction. In general we find a
qualitative agreement. In particular the frequency shift in the different
formalisms has the same scaling with the parameters (density and wavenumber)
both for ion-acoustic and Langmuir waves. However, there is a discrepancy
concerning a numerical factor. This is discussed in more detail in the final
section of the manuscript.

\section{The exchange correction at $T = \SI{0}{K}$}

In a previous paper, Ref.\ \cite{Zamanian-exchange}, the exchange
contribution to the evolution equation of the Wigner function was derived
(see Eq. (7) of Ref.\ \cite{Zamanian-exchange}). The correction was obtained
by writing down the first equation in the BBGKY-hierarchy and writing the
two-particle density matrix as a anti-symmetric product of one-particle
density matrices. The treatment was here limited to electrostatic fields.
For a generalization allowing for electromagnetic fields, see Ref. \cite%
{Zamanian-II-exchange}. Eq. (7) of Ref.\ \cite{Zamanian-exchange} was
further simplified by considering a plasma without spin polarization and
summing over all spin states, and also by taking the long scale limit (where
the macroscopic scale length is assumed to be much longer than the de
Broglie wavelength). The long scale assumption implies that the Wigner
function reduces to the Vlasov limit, in which case the Wigner function
becomes similar to a classical distribution function. This means that the
evolution equation reduces to the Vlasov equation with a correction term due
to the exchange effects. The resulting expression (Eq. (12) of Ref. \cite%
{Zamanian-exchange}) with the exchange correction written in the right hand
side reads 
\begin{widetext} 
\begin{align} 
	\partial_t f (\mathbf x, \mathbf p, t) 
	+ \frac{\mathbf p}{m} \cdot \nabla_x f(\mathbf x, \mathbf p, t) 
	+ e \mathbf E (\mathbf x,t) \cdot \nabla_p f(\mathbf x ,\mathbf p, t) 
	 =  \,
	 \frac{1}{2} \partial_p^i \int d^3\! r \, d^3\! q \,\, e^{ - i \mathbf r \cdot \mathbf q / \hbar} 
	[\partial_r^i V (\mathbf r)] 
	f \left( \mathbf x - \frac{\mathbf r}{2} , \mathbf p + \frac{\mathbf q}{2}, t \right) 
	f \left( \mathbf x - \frac{\mathbf r}{2} , \mathbf p - \frac{\mathbf q}{2}, t \right)  
\notag \\ 
	- \frac{i \hbar}{8}  
	\partial_p^i \partial_p^j \cdot \int d^3\! r \, d^3\! q \,\, e^{ - i \mathbf r \cdot \mathbf q / \hbar} 
	[\partial_r^i V (\mathbf r) ]
	\left[ f \left( \mathbf x - \frac{\mathbf r}{2} , \mathbf p - \frac{\mathbf q}{2},  t \right) 
	\left( \overleftarrow \partial_x^j  - \overrightarrow \partial_x^j \right)
	f \left( \mathbf x - \frac{\mathbf r}{2} , \mathbf p + \frac{\mathbf q}{2} , t \right)  \right]
\label{23}  
\end{align} 
\end{widetext}Here $f$ denotes the electron distribution function, $\mathbf{E%
}$ is the electrical field, $-e~\ $and $m_{e}$ is the electron charge and
mass respectively, $h=2\pi \hbar $ is Planck's constant and we use $\mathbf{x%
}$ and $\mathbf{r}$ for position vectors and $\mathbf{p}$ and $\mathbf{q}$
for momentum vectors. Furthermore, $\partial _{x}^{i}\equiv \partial
/\partial x_{i}$ and analogously for $\partial _{p}^{i}$ and $\partial
_{r}^{i}$. An arrow above an operator indicates in which direction it acts.
We have also used the summation convention so that a sum over indices
occurring twice in a term is understood. Finally $V(\mathbf{r})=$ $%
e^{2}/4\pi \varepsilon _{0}\left\vert \mathbf{r}\right\vert $ is the Coulomb
potential. Here we will use Eq. \eqref{23} to consider linear wave
propagation treating the exchange term on the right hand side
perturbatively. Thus we linearize Eq. \eqref{23} and make a plane wave
ansatz $E_{z}=E\exp (-i\omega t+ikz)$, $f(\mathbf{x},\mathbf{p}%
,t)=f_{0}(p)+f_{1}\exp (-i\omega t+ikz)$.

To first order in a long wavelength expansion, the linearized version of Eq. %
\eqref{23} then reduces to 
\begin{widetext}
\begin{equation} 
	- i \left(  \omega - \frac{k p_z}{m_e} \right) f_1 (\mathbf p) 
	 = 
	- q E \partial_{p_z} f_0 (p)
	+ \frac{2 i \hbar^2 k q^2}{ \epsilon_0}  \nabla_p \cdot     
	\int d^3 q  
	\frac{\mathbf q  }{ \left| \mathbf q \right|^2 } 
	\left[
	\partial_{p_z} f_0 \left( \left| \mathbf p + \mathbf q  \right| \right)
	f_1 \left( \mathbf p - \mathbf q \right) 
	\right] . \label{eq:vlasovExchange}
\end{equation}
\end{widetext}
Previously the correction to the dispersion relations for ion acoustic waves
(see Ref. \cite{Zamanian-exchange}) and for Langmuir waves (see Ref. \cite%
{Zamanian-II-exchange}) has been found considering a Maxwellian background,
by making the approximation $f_{1}\approx \tilde{f}_{1}$ in the exchange
term. The symbol $\tilde{f}_{1}=qE/[i(\omega -kp_{z}/m_{e})]\partial
f_{0}/\partial p_{z}$ denotes the solution when exchange effects are
neglected.

Here we will study the same problems but for the case when $T=\SI{0}{K}$. As
we will demonstrate the problem is analytically tractable owing to the
simple form of the background distribution function $f_{0}$ which is now
given by 
\begin{equation}
f_{0}(\mathbf{p})=\left\{ 
\begin{array}{cc}
2/(2\pi \hbar )^{3}, & \quad \quad \left\vert \mathbf{p}\right\vert \leq p_{%
\text{F}} \\ 
0, & \quad \quad \left\vert \mathbf{p}\right\vert >p_{\text{F}}\ ,%
\end{array}%
\right.  \label{FD-distribution}
\end{equation}%
where $p_{\text{F}}\ =\hbar (3\pi ^{2}n_{0})^{1/3}$ is the Fermi momentum
(we will also use the notation $v_{\text{F}}=p_{\text{F}}/m_{e}$ for the
electron Fermi velocity below), and $n_{0}$ is the equilibrium electron
number density. In particular, taking $f_{1}=\tilde{f}_{1}$ in the exchange
term, combining Eqs. (2) and (\ref{FD-distribution}) with Poisson's equation
and computing the ion charge density classically we obtain%
\begin{equation}
1=\chi _{0}+\chi _{\text{1}},
\end{equation}%
where 
\begin{equation}
\chi _{0}=\frac{\omega _{i}^{2}}{\omega ^{2}}+\frac{3\omega _{e}^{2}}{%
2k^{2}v_{\text{F}}^{2}}\int_{-1}^{1}dz\frac{z}{\omega /(kv_{\text{F}})-z}
\label{susceptibility}
\end{equation}%
is the combined ion and electron susceptibility in the absence of exchange
effects. Here we have neglected a contribution from a small but finite ion
Fermi velocity (i.e. we have used that $v_{\text{Fi}}\ll \omega /k$). In the
expression above $\omega _{e}$ and $\omega _{i}$ denote the electron and ion
plasma frequency respectively. The exchange correction is given by 
\begin{equation}
\chi _{1}=-\frac{q^{4}k}{8\pi ^{6}\epsilon _{0}^{2}\hbar ^{4}m_{e}}\int d^{3}%
\mathbf{p}\,d^{3}\mathbf{q}\,\delta (|\mathbf{p}-\mathbf{q}|-p_{\text{F}%
})\delta (|\mathbf{p}+\mathbf{q}|-p_{\text{F}})F
\end{equation}%
with 
\begin{equation}
F=\frac{1}{(\omega -kp_{z}/m_{e})^{2}}\frac{1}{\omega -k(p_{z}-q_{z})/m_{e}}%
\frac{p_{z}}{|\mathbf{p}|^{2}}.
\end{equation}%
By changing integration variables to $\mathbf{u}_{1}=\mathbf{p}+\mathbf{q}$, 
$\mathbf{u}_{2}=\mathbf{p}-\mathbf{q}$, with the Jacobian equal to $1/8$, we
can use the properties of the Dirac delta functions to reduce the problem to
an integral over two spheres. Introducing spherical coordinates $\mathbf{u}%
_{i}=u_{i}(\cos \varphi _{i}\sin \theta _{i},\sin \varphi _{i}\sin \theta
_{i},\cos \theta _{i})$ we explicitly get $\chi _{1}=m_{e}^{2}q^{4}I^{\prime
}$ $/8\pi ^{6}\epsilon _{0}^{2}k^{2}\hbar ^{4}$with $I^{\prime }$ given by 
\begin{widetext}\begin{equation}
	I' = \frac{1}{4} \int \frac{\cos\theta_1 \cos\theta_2}{( 2- 2 \cos\psi)} \frac{\cos\theta_1 - \cos\theta_2}{\alpha - \cos\theta_2 } \frac{d(\cos\theta_1)d(\cos\theta_2) d \varphi_1 d\varphi_2 }{(\alpha - (\cos \theta_1 + \cos\theta_2)/2)^2}
\end{equation}
\end{widetext}where $\psi $ is the angle between $\mathbf{u}_{1}$ and $%
\mathbf{u}_{2}$ and $\alpha =\omega /(kv_{\text{F}})$ is the dimensionless
phase velocity. By the spherical law of cosines $\cos \psi =\cos \theta
_{1}\cos \theta _{2}+\sin \theta _{1}\sin \theta _{2}\cos (\varphi
_{1}-\varphi _{2})$. By changing variables to $\varphi _{1},\tilde{\varphi}%
=\varphi _{1}-\varphi _{2}$, the $\varphi _{1}$ integral is trivial. The $%
\tilde{\varphi}$ integral can then be performed, giving a factor $\pi
/(a^{2}-b^{2})^{1/2}$ where $a=1-\cos \theta _{1}\cos \theta _{2}$ and $%
b=\sin \theta _{1}\sin \theta _{2}$. Since $a^{2}-b^{2}=(\cos \theta
_{1}-\cos \theta _{2})^{2}$, we have that $I^{\prime }=-2^{-1}\pi ^{2}I$
where $I$ is the double integral 
\begin{equation}
I=\int_{-1}^{1}dx\int_{-1}^{1}dy\,\frac{xy}{\alpha -y}\frac{\sgn(x-y)}{%
(\alpha -(x+y)/2)^{2}}.  \label{eq:exchangeIntegral}
\end{equation}%
%
%
%
%
%
%
%
%
%
%
%
%
%
%
%
%
%
%
%
%
%
%
%
%
At this point, no approximations have been made beyond those leading to the
evolution equation~\eqref{eq:vlasovExchange}, taking $f_{1}=\tilde{f}_{1}$,
and $T=\SI{0}{K}$. For these approximations the exchange correction is
therefore proportional to \eqref{eq:exchangeIntegral} for all frequency
regimes. Thus Poisson's equation, with exchange corrections, is 
\begin{equation}
1=\chi _{0}-\frac{9\omega _{e}^{4}\hbar ^{2}}{16k^{2}m_{e}^{2}v_{\text{F}%
}^{6}}I.  \label{9}
\end{equation}%
where $\chi _{0}$ is given by Eq. (\ref{susceptibility}). From now on we
must treat the low-frequency ion-acoustic case separately from the
high-frequency Langmuir case in order to simplify the expression for $I$.

\paragraph{Ion-acoustic waves}

The integral $I$ can be evaluated analytically in terms of $\alpha $ with
computer algebra software. We will consider the quasi-neutral limit ($\omega
\ll \omega _{i}$) where the left hand side of Eq.\ \eqref{9} is negligible.
For ion-acoustic waves we can then make the approximation that $\alpha =%
\sqrt{m_{e}/(3m_{i})}$ when solving the integral \eqref{eq:exchangeIntegral}%
. 
It is also possible to plug in the value of $\alpha $ and evaluate the
integral numerically. The dispersion relation for ion-acoustic waves with
exchange corrections is then 
\begin{equation}
\omega ^{2}=\alpha ^{2}k^{2}v_{\text{F}}^{2}\left[ 1-\frac{\hbar ^{2}\omega
_{e}^{2}}{3m_{e}^{2}v_{\text{F}}^{4}}(14.9+7.11i)\right]
\label{Ion-acoustic-DR}
\end{equation}%
To the best of our knowledge this result has not been derived before.

\paragraph{Langmuir waves}

In the high-frequency regime $\omega /k\gg v_{\text{F}}$, an expansion of $I$
in powers of $v_{\text{F}}\ k/\omega $ can be made. With the lowest order
non-vanishing correction, the dispersion relation in this regime is 
\begin{equation}
\omega ^{2}=\omega _{e}^{2}+\frac{3}{5}v_{\text{F}}^{2}k^{2}-\frac{3\hbar
^{2}\omega _{e}^{2}k^{2}}{20m_{e}^{2}v_{\text{F}}^{2}}.  \label{Langmuir-DR}
\end{equation}%
This is in exact agreement with previous calculations using several
different methods, see Refs. \cite{Roos-1961,Kanazawa-1960,Nozieres-1958}.

\section{DFT comparison}

Eq. (\ref{eq:exchangeIntegral}) together with the specific results (\ref%
{Ion-acoustic-DR}) and (\ref{Langmuir-DR}) are the main results of the
present paper. The exact agreement of (\ref{Langmuir-DR}) with previous
results also confirms the validity of (\ref{eq:exchangeIntegral}). A
strength of the quantum kinetic formalism is that it follows from first
principles, and that it can address wave particle interaction, such as the
enhanced Landau damping rate found in Eq. (\ref{Ion-acoustic-DR}). However,
a drawback is that the formalism is difficult to use for more complicated
problems. \ Thus there is a need to develop theories that are easier to
apply in a more general context. One such possibility is offered by DFT. A
commonly used exchange potential (see e.g. Refs. \cite%
{Manfredi-DFT,DFT-HF-LF,DFT-LF,DFT-semicond}) computed from DFT is 
\begin{equation}
V_{x}=\frac{0.985(3\pi ^{2})^{2/3}}{4\pi }\frac{\hbar ^{2}\omega _{e}^{2}}{%
m_{e}v_{F}^{2}}\left( \frac{n}{n_{0}}\right) ^{1/3}  \label{DFT-potential}
\end{equation}%
that has been used for low-frequency ion-acoustic phenomena (e.g. Refs. \cite%
{DFT-HF-LF,DFT-LF} ) as well as high-frequency Langmuir waves (e.g. Refs. 
\cite{Manfredi-DFT,DFT-HF-LF}). Including the contribution from Eq. (\ref%
{DFT-potential}) in the electron momentum equation (see e.g. Ref. \cite%
{Manfredi-DFT}) and treating the term as a small perturbation when
calculating the ion-acoustic dispersion relation, we get a qualitative
agreement with Eq. (\ref{Ion-acoustic-DR}). In particular the ion-acoustic
frequency is decreased and this change scale as $\hbar ^{2}\omega
_{e}^{2}/m_{e}^{2}v_{F}^{4}\,$, in accordance with (\ref{Ion-acoustic-DR}).
The numerical value of the frequency shift deviates rather significantly
from our result, however. We note that if we make the adjustment $%
0.985\rightarrow 6.52$ of the numerical pre-factor in (\ref{DFT-potential})
we would get agreement with the real part of the frequency in Eq. (\ref%
{Ion-acoustic-DR}). Here we stress that the wave damping cannot be compared
with the DFT formalism, as this requires a quantum kinetic framework.

The same comparison can be made in the high-frequency regime. Thus we again
include the contribution from (\ref{DFT-potential}) in the electron momentum
equation, calculating the frequency shift of the Langmuir dispersion
relation. Also here we have qualitative agreement, i.e. the Langmuir
frequency increases proportional to $k^{2}$, and the scaling with density is
in accordance with the factor $\hbar ^{2}\omega _{e}^{2}/m_{e}^{2}v_{F}^{4}.$
This time the numerical accuracy is better, although not perfect, and we
need to make a less significant substitution of the numerical factor in Eq. (%
\ref{DFT-potential}), $0.985\rightarrow 1.23$ in order to get agreement with
our result (\ref{Langmuir-DR}). Thus we conclude that the DFT exchange
potential Eq. (\ref{DFT-potential}) are in qualitative agreement with the
results obtained here, but that the numerical accuracy is better for
Langmuir waves than for ion-acoustic waves.

\section{Summary and conclusion}

In the present paper we have computed the exchange contribution to the
ion-acoustic dispersion relation in a plasma, using a quantum kinetic
formalism. The validity of our approach have been confirmed by comparison
with similar results for high-frequency Langmuir waves \cite%
{Roos-1961,Kanazawa-1960}, in which case we get exact agreement for the
exchange correction. This is to be expected, as our formalism as well as
that of Refs. \cite{Roos-1961,Kanazawa-1960} is based on first principles.
While the quantum kinetic formalism is of fundamental importance, it has the
drawback of producing complicated formulas (cf. Eq. (1)) that can only be
solved perturbatively. In fact even a perturbative treatment is far from
straightforward. Thus there is a need for a formalism that can be used in
more complex situations. Such a possibility is offered by DFT, where the
resulting exchange potentials can be used in fluid theories, and
straightforwardly applied to \ a number of linear and nonlinear problems 
\cite{Manfredi-DFT,DFT-HF-LF,DFT-LF,DFT-semicond}. However, as the
computation of DFT potentials typically involves approximations (e.g. the
local density approximation (LDA)) whose accuracy is unknown, there is a
need to evaluate DFT potentials against independent methods. A key
motivation in the present paper has been to evaluate the DFT potentials used
in Refs. \cite{Manfredi-DFT,DFT-HF-LF,DFT-LF,DFT-semicond} against the
quantum kinetic results computed in our formalism. This comparison reveals
that there is a reasonable qualitative agreement. In both cases the relative
magnitude of the exchange term scales as $\hbar ^{2}\omega
_{e}^{2}/m_{e}^{2}v_{F}^{4}$, and both the Langmuir and ion-acoustic wave
frequencies decreases due to the exchange interaction, in agreement with
Eqs. (\ref{Ion-acoustic-DR}) and (\ref{Langmuir-DR}). The numerical value of
this frequency shift differ somewhat, however. To some extent this can be
fixed by replacing the numerical pre-factor in the exchange potential. If
this approach is chosen, we note that different substitutions must be used
for low-frequency and high-frequency phenomena (i.e. $0.985\rightarrow 6.52$
in the former case, $0.985\rightarrow 1.23$ in the latter case). It is an
open question to what extent this result is robust (i.e. whether the same
numerical coefficient is a good approximation for different problems at a
given frequency scale), or if more advanced expressions for the exchange
potentials are needed to cover a broad spectrum of problems.

The fact that there is not a perfect agreement between the DFT exchange
potential and the quantum kinetic theory should not be overly surprising.
The DFT potentials used in Refs. \cite%
{Manfredi-DFT,DFT-HF-LF,DFT-LF,DFT-semicond} have not necessarily been
optimized for the situation we have been studying. In our case the fields
are dynamically varying (such that time-dependent density functional theory
(TDDFT) applies), and the system is weakly collisional such that the
collision-free Vlasov equation holds to leading order. The systems of
relevance for the present study generally include plasmas with a high
density and a modest temperature $T\ll T_{F}$. In a laboratory context this
applies to e.g. solid state plasmas and inertial confinement fusion plasmas
before the heating stage, and in an astrophysical context this include e.g.
white dwarf stars. In that case the perturbed distribution function will be
comparatively far from thermodynamic equilibrium, and approaches which do
not include the full quantum kinetic details run the risk of losing
information. Nevertheless an approach based on density functional theory is
a valuable option, in particular for problems where the approach used here
becomes too complicated.

\end{document}